\documentclass[12pt,preprint]{aastex}

\newcommand{\Msun}{$M_{\odot}$}

\def\h2{H$_{2}$}

\input epsf

\shorttitle{Effects of Pop III Stars on Reionization}
\shortauthors{Venkatesan, Tumlinson, \& Shull}

\begin{document}

\title{Evolving Spectra of Pop III Stars: \\ Consequences for
  Cosmological Reionization} 
\author{Aparna Venkatesan\altaffilmark{1}, Jason Tumlinson\altaffilmark{2}, \& J. Michael Shull\altaffilmark{3}} 
\affil{Center for Astrophysics and Space Astronomy, \\ Department of
  Astrophysical and Planetary Sciences, \\ UCB 389, University of Colorado,
  Boulder, CO 80309-0389}
\altaffiltext{1}{NSF Astronomy and Astrophysics Postdoctoral Fellow}
\altaffiltext{2}{Present address: Department of Astronomy and Astrophysics,
  University of Chicago, Chicago, IL 60637}
\altaffiltext{3}{Also at JILA, University of Colorado and National
Institute of Standards and Technology}
\email{(aparna, tumlinso, mshull)@casa.colorado.edu}

\begin{abstract}

We examine the significance of the first metal-free stars (Pop III) for the
cosmological reionization of H~I and He~II.  These stars have unusually
hard spectra, with the integrated ionizing photon rates from a Pop III
stellar cluster for H~I and He~II being 1.6 and $10^5$ times stronger
respectively than those from a Pop II cluster.  For the currently favored
cosmology, we find that Pop III stars alone can reionize H~I and He~II at
redshifts, $z \simeq$ 9 (4.7) and 5.1 (0.7) for continuous (instantaneous)
modes of star formation.  More realistic scenarios involving combinations
of Pop III and Pop II stellar spectra yield similar results for
hydrogen. Helium never reionizes completely in these cases; the ionization
fraction of He~III reaches a maximum of about 60\% at $z \sim 5.6$ if Pop
III star formation lasts for $10^9$ yr. Future data on H~I reionization can
test the amount of small-scale power available to the formation of the
first objects, and provide a constraint on values of $\sigma_8$ $\la$ 0.7.
Since current UV observations indicate an epoch of reionization for He~II
at $z \sim 3$, He~II may reionize more than once.  Measurements of the
He~II Gunn-Peterson effect in the intergalactic medium at redshifts $z \ga$
3 may reveal the significance of Pop III stars for He~II reionization,
particularly in void regions that may contain relic ionization from early
Pop III stellar activity.

\end{abstract}

\keywords{cosmology: theory---intergalactic medium}

\section{Introduction}

The nature, formation sites, and epochs of the first stars in the universe
are some of cosmology's most intriguing yet unresolved questions today.
Theoretical studies of the effects of these objects on the high-redshift
intergalactic medium (IGM) and on galaxy formation have a rapidly expanding
literature, driven in part by the potential to test the predictions from
such theories with data in the near future. Recent work on the first stars
has focused on signatures such as the effects of stellar radiation and
nucleosynthesis on their host galaxies and the IGM \citep{go97,hl97,
ferrara00, abia, rgshull02} -- loosely grouped under stellar and supernova
(SN) feedback -- and the potential presence of large numbers of stellar
remnants in galactic halos in baryonic dark matter scenarios
\citep{fields}.  These studies were directly motivated by observations of
the reionization \citep{becker, kriss} and trace metal enrichment
(\citealt{song01}, 2002) of the high-$z$ IGM, and the detection of
solar-mass dark objects in our Galactic halo by microlensing experiments
\citep{alcock}. Although it is unclear if the same population of early
stars can be tied unambiguously to all of these data, it is likely that
they contributed significantly to the ionizing photon budget and metal
production at early times.

Beyond such signatures, there has also been considerable interest in the
typical masses of the first stars and the preferred environments, if any,
in which they form. There is no theoretical basis on which one can {\it a
priori} rule out a stellar initial mass function (IMF) that was different
in the past. Indeed, arguments for a primordial IMF biased towards higher
masses have been proposed for some time \citep{carr84,larson,
abel,herferr,bromm,naka}, although there is no indication for an
environment-dependent IMF from data of a variety of local star-forming
regions \citep{kroupa}. As for the nature of the typical galaxy that hosts
the first stars, this depends critically on the availability of coolants
within virialized halos so that the fragmentation necessary for star
formation may commence. Several authors \citep{teg97,ciardi} have argued
that modest levels of early stellar activity can generate sufficient
far-ultraviolet radiation in the Lyman-Werner bands (11.2 -- 13.6 eV) to
photodissociate all of the remaining H$_2$ in the universe, well before the
associated H~I ionizing flux has built up to values sufficient for
reionization. A long pause in global star formation would then ensue, owing
to ``negative feedback'', and would resume only when halos of virial
temperature $\ga$ 10$^4$ K collapse, corresponding to the threshold for the
onset of H line cooling.  One way to overcome such negative feedback might
be through the presence of X-rays or ionizing UV photons from the first
luminous sources, which could boost the free electron fraction and hence
the amount of H$^{-}$-catalyzed H$_2$, leading to a compensatory positive
feedback \citep{rgshull01}. It remains unclear how effective this is in
overcoming the negative feedback from infrared and Lyman-Werner band
photons \citep{har,venkgs01}. There is also the possibility that sufficient
metals are injected into the interstellar medium (ISM) soon after the very
first stars form. In this case, the distinction between halos cooling by
H$_2$ versus H becomes irrelevant, assuming that the metals can be retained
in the cold star-forming gaseous component within individual halos. This
problem remains unresolved currently, but it is clear that the chemistry of
high-$z$ halos is critical to when and where the first stars formed.

{\it A priori}, we would expect early generations of stars forming from
primordial gas to be metal-free in composition, although no surviving
members of such populations have been detected to date.  Recent studies of
stars of zero metallicity, $Z$ (\citealt{tumshull,bromm, cojazzi};
\citealt{schaerer}, and references therein), which we henceforth refer to
as Pop III, have demonstrated that $Z = 0$ stars are fundamentally
different in nature and evolutionary properties from their low-$Z$
counterparts. In particular, \citet{tumshull} showed, in a calculation of
the zero-age main sequence of these stars, that their harder ionizing
spectra could be relevant for both the H~I and He~II reionization of the
IGM. This work has subsequently been extended to full calculations of the
evolving spectra of Pop III stellar populations in \citet{tsv02}
(henceforth Paper I), where we found that, for a Salpeter IMF, the
integrated ionizing photon rate from a Pop III cluster for H~I and He~II is
respectively 1.6 and $10^5$ times stronger than that for a $Z = 0.001$
cluster \citep{starburst} of the same mass\footnote{In this comparison, we
do not include the contribution of Wolf-Rayet stars which can boost the
ionizing radiation from a Pop~II stellar cluster. In fact, the existence of
the Wolf-Rayet phase in $Z=0$ stars is questionable, as discussed in Paper
I, given that these objects are unlikely to experience strong mass loss.
We are primarily interested here in a direct comparison of the ionizing
radiation from the main-sequence phases of Pops III and II.}.  In Paper I,
we examined the evolving spectra of Pop III stars and their observational
signatures such as broad-band colors and emission lines. In this paper,
which is intended as a companion work to Paper I, we focus on the
significance of such metal-free stellar populations for cosmological
reionization, under the assumption that they form in a present-day IMF.

We present a brief review here of the status of theoretical models and data
on reionization.  Spectroscopic studies of high-$z$ quasars and galaxies
blueward of their rest-frame H~I and He~II Ly$\alpha$ emission have
revealed that He~II reionization occurs at $z \sim$ 3 \citep{kriss} and
that of H~I before $z \sim 6$ \citep{becker}. Such spectroscopic
observations, along with increasingly precise data on the cosmic microwave
background (CMB), are beginning to place strong complementary bounds on the
redshift of H~I reionization, $z_{\rm reion, H}$. At one end, current CMB
data on the temperature anisotropy at degree and sub-degree scales provide
an upper limit of about 0.3 for the electron-scattering optical depth to
reionization, which may be translated into a model-dependent constraint of
$z_{\rm reion, H}$ $\la$ 25 \citep{wangtz01}.  Ongoing and future CMB
observations (see
http://www.hep.upenn.edu/$\sim$max/cmb/\newline experiments.html, and
http://background.uchicago.edu/$\sim$whu/cmbex.html for links to various
CMB experiments) will provide improved constraints on $z_{\rm reion, H}$
through the detection of polarization in the CMB at large angular scales
\citep{staggs}. At the other end of the range for $z_{\rm reion, H}$, the
IGM appears to be highly ionized up to $z \sim 6$
\citep{fan1,dey}. \citet{becker} recently detected the H~I Gunn-Peterson
(GP) trough in the spectrum of the highest-redshift quasar known to date at
$z = 6.28$ \citep{fan3}, which may indicate that H~I reionization occurs
not far beyond $z \sim 6$.  This claim has been challenged, however, by the
subsequent observation of Ly$\alpha$ emission in a $z = 6.56$ galaxy
\citep{hu02}.  The extent to which this Ly$\alpha$ emission line has been
eroded by the damping wing of Ly$\alpha$ absorption in the IGM
\citep{escude} is, however, unclear.  It is therefore difficult to assess
whether H~I reionization is complete at $z \sim$ 6.5 based on this one
object.  Although the detection of the GP trough in a single line of sight
is not definitive evidence of the global reionization of the IGM, it may
probe the end of the gradual process of inhomogeneous reionization,
coinciding with the disappearance of the last neutral regions in the
high-$z$ IGM.  Reionization of H~I at $z$ $\sim$ 6 -- 9 would still be
consistent with the lower end of the range of redshifts, $z \sim$ 6 -- 20,
predicted by theoretical models, both semi-analytic
\citep{tsb94,gs96,hl97,valsilk,mhrees99,mirees00} and based on numerical
simulations \citep{cenost93,gnedin00,ciardi,benson2}.

The reionization of the IGM subsequent to recombination at $z \sim$ 1000 is
thought to have been caused by increasing numbers of the first luminous
sources. Although there are a variety of models for the astrophysical
objects or processes that could have reionized the IGM, the leading
scenarios involve photoionization by sources with soft or hard ionizing
spectra, or equivalently stellar-type or QSO-type models
respectively. Clearly, this division of source populations according to
their spectral properties is no longer valid if the first stars generated
hard ionizing radiation by virtue of their metal-free composition (Paper I)
or if they formed in an IMF biased towards extremely high masses
\citep{bromm}.  The large majority of currently favored reionization models
involve stars rather than QSOs, for a number of observationally motivated
reasons (see \citealt{venk02} for a detailed discussion on this
point). These reasons include the apparent decrease in the space density of
large, optically bright quasars up to $z \sim$ 6.3 beyond a peak at $z
\sim$ 3 \citep{fan3,shaver}, so that their UV and X-ray photons are
insufficient for H~I reionization \citep{mad99,venkgs01}. However, the
nature of the reionizing sources is highly uncertain at present, and one
can neither confirm nor exclude stars, faint QSOs (``mini-QSOs''), or a
combination of high-$z$ source populations. In this paper, we are not
specifically advocating that Pop III stars are solely responsible for
reionization. Rather, our main goal is to examine the consequences of an
epoch of metal-free star formation for the H~I and He~II reionization of
the IGM, given a model of reionization with currently favored values for
the input parameters.

We organize the paper as follows. In \S 2, we present the newly calculated
evolving spectra of Pop III from Paper I, and we describe the reionization
model used in this work. In \S 3, we present our results on the effects of
Pop III and Pop II stars on the H~I and He~II reionization of the IGM for a
number of potential scenarios of high-$z$ star formation. We also discuss
the potential constraint offered by the reionization epoch on the amount of
small-scale power in structure formation models, and we speculate on the
fate of partially or fully ionized He~III in underdense regions of the IGM
whose detection lies on the threshold of current capabilities.  We present
our conclusions in \S 4.

\section{The Reionization Model}

We use the semi-analytic stellar reionization model described in Venkatesan
(2000; henceforth V2000), and consider only the effects of stars. This
model calculates the fraction of baryons in collapsed halos with the
Press-Schechter formulation, and, given a prescription for star formation
and the generated ionizing radiation, tracks the hydrogen and helium
reionization of the IGM. We adopt the methods developed in V2000, replacing
the standard cold dark matter (SCDM) model of V2000 with a $\Lambda$CDM
cosmology, and we significantly improve the solution for the growth of
ionization regions around individual halos, as described below.

We take the primordial matter power spectrum of density fluctuations to be,
$P(k) \propto k^n$ $T^2 (k)$, where $n$ is the index of the scalar power
spectrum, and the matter transfer function $T(k)$ is taken from
\citet{ehu98}. We assume that there are no tensor contributions to
$P(k)$. We normalize $P(k)$ to the present-day rms density contrast,
$\sigma_8$, over spheres of radius 8 $h^{-1}$ Mpc, in which we assume that
the bias factor is unity (i.e., that light traces the underlying mass
distribution). We choose this normalization rather than the COBE
normalization from the CMB, as the physical scales associated with
$\sigma_8$ are relatively close to those relevant for reionization and the
formation of the first luminous objects.

We track the fraction, $F_{\rm B}$, of all baryons in collapsed dark matter
halos by the Press-Schechter formalism, allowing star formation only in
massive halos of virial temperature $\ga$ 10$^4$ K, corresponding to the
mass threshold for the onset of hydrogen line cooling. In the absence of
metals, H is likely to be the primary coolant, since H$_2$ is easily
destroyed by trace levels of stellar UV radiation. In the case of low-mass
halos, the interplay between positive and negative feedback on H$_2$
formation and destruction can lead to time-varying H$_2$ abundances. This
in turn causes episodic star formation, so that the generated ionization
fronts (I-fronts) remain trapped in the host galaxy and do not have a large
effect on the IGM \citep{rgshull02}. We therefore focus on reionization
from high-mass halos.

We assume that the fraction of baryons in each galaxy halo forming stars is
given by $f_\star$, and that the fraction of H~I and He~II ionizing photons
escaping from individual halos are given respectively by $f^{\rm H}_{\rm
esc}$ and $f^{\rm He}_{\rm esc}$. Our assumed values for these parameters
are discussed below.  We assume that the ionizing photons propagate
isotropically from their host galaxies into the IGM, generating an ionized
sphere around each source of radius $r_i$.  We solve for the size of the
ionized regions associated with each such star-forming galaxy, which, when
integrated over all halos, yields at each redshift the volume filling
factor of H~II ($F_{\rm H II}$) or He~III ($F_{\rm He III}$). Reionization
is defined as the epoch when individual ionized regions overlap, when
$F_{\rm H II}$ = 1 or $F_{\rm He III}$ = 1; we return to this below.

Since most cosmological parameters are becoming increasingly
well-constrained by combined CMB, large-scale structure, and high-$z$ SN Ia
observations, the remaining uncertainty in the ``initial conditions'' for
reionization rests largely with $\sigma_8$, and to a lesser degree on
$n$. The scalar power spectrum index is currently measured by the above
techniques to about 10\% error, $n \sim$ 0.9 -- 1.1. The data on $\sigma_8$
are currently divided between two values, one at about 0.9 -- 1.0
\citep{evrard,refregier}, and the other at a substantially lower value of
about 0.7 \citep{reip}. We choose $\sigma_8$ = 0.9 in this work as a
representative intermediate value. We caution that $\sigma_8$ can be
reduced in value only to a certain degree in reionization models;
otherwise, there is a significant loss of power on small scales, which may
lead to reionization too late to be consistent with current data. Some
leverage can be regained by increasing $n$ or by fine-tuning the
astrophysical parameters in the model, but this may not prove sufficient.
We return to this topic in \S 3.3.

We set $f_\star$ = 0.05 for all galaxy halos. This value is consistent with
the findings of numerical simulations of star formation in early halos and
with the constraint of avoiding the overenrichment of the IGM in metals by
$z \sim 3$ (see V2000 and references therein), for both continuous and
bursting modes of star formation.  We set $f^{\rm H}_{\rm esc}$ = 0.05
\citep{dehar,dsf,leitherer}, to be consistent with data from the local
universe, particularly of high-mass systems, and assuming that local
systems describe the conditions found in the first star-forming galaxies.
Although observations of Lyman-continuum emission from Lyman-break galaxies
at $z \sim$ 3.4 by \citet{steidel} indicate values of $f^{\rm H}_{\rm esc}$
exceeding 0.5, it is not clear how representative these systems are of
high-$z$ star formation, and how much these high values are an artifact of
the observational procedure.

Assigning a value to $f^{\rm He}_{\rm esc}$ for individual galaxies is
somewhat more uncertain. Given the greater recombination rate of He~III
relative to H~II and the fact that most astrophysical sources have much
lower He~II ionizing fluxes relative to the values for H~I, one would
expect $f^{\rm He}_{\rm esc}$ to be lower than $f^{\rm H}_{\rm esc}$. On
the other hand, a combination of the first wave of ionizing photons,
combined with the clearing of the ISM in high-$z$ galaxies by the first
SNe, may provide ``equal-opportunity chimneys'' for the escape of H~I and
He~II ionizing photons. Given the lack of data on $f^{\rm He}_{\rm esc}$,
and the above factors, we set $f^{\rm He}_{\rm esc}$ = 0.5~$\times$~$f^{\rm
H}_{\rm esc}$ = 0.025, as a reasonable first guess at this quantity;
certainly, we do not expect $f^{\rm He}_{\rm esc}$ to exceed $f^{\rm
H}_{\rm esc}$. Although we have adopted reasonable values for these
astrophysical parameters, they are likely to have some dependence on
redshift and individual galaxy masses; thus, the assumption of a constant
value for them is an oversimplification. However, the relevant issue is
that they have to combine in such a way so as to be consistent with the
above constraints, and lead, for most reionization models, to H~II and
He~III reionization at epochs consistent with those from observations.

We include the effects of inhomogeneity in the IGM through a clumping
factor, $c_L$, rather than assuming a smooth IGM as in V2000. We define
$c_L$ to be the space-averaged clumping factor of photoionized hydrogen or
helium, $c_L$ $\equiv$ $\langle n_i^2 \rangle$/$\langle n_i \rangle^2$,
where $i$ corresponds to H~II or He~III.  Here, we take $c_L$ to be the
same for both H~II and He~III. The parameter $c_L$ can affect the epoch of
reionization significantly, with higher values leading to delayed
reionization. However, within a specific framework of cosmology and
structure evolution, $c_L$ is a derived rather than an independent
parameter. Although $c_L$ is necessarily redshift-dependent, we assume
$c_L$ = 30, which is a reasonable average from, e.g., \citet{mhrees99} and
\citet{go97}, for the redshifts we consider here.  This value of $c_L$ also
results in reionization epochs for H~I and He~II that span ranges
consistent with observations, for the various cases that we consider in \S
3.

As an aside, we note that $c_L$ and $f_{\rm esc}$ are inherently related,
owing to their strong dependence on the scale on which they are defined.
Some numerical simulations of reionization by stars appear to require
values of $f^{\rm H}_{\rm esc}$ $\sim$ 0.5 -- 1 \citep{gnedin00,benson2} in
order to have the calculated values of $z_{\rm reion, H}$ be consistent
with observations. In the case of \citet{gnedin00}, this seemingly high
value of $f^{\rm H}_{\rm esc}$ is misleading, as it is in fact defined as
the value at the surface of the star and does not correspond to the $f^{\rm
H}_{\rm esc}$ relevant for those ionizing photons that reach the IGM, as
defined in this work. The loss of ionizing photons within individual halos
is compensated through a clumping factor for H~II regions that accounts for
this, so that at early times, the $c_{\rm HII}$ in \citet{gnedin00} exceeds
values of a few thousand. For models that define $f^{\rm H}_{\rm esc}$ at
the halo/IGM scale, the appropriate $c_{\rm HII}$ is the one in
\citet{go97} or \citet{mhrees99} rather than in \citet{gnedin00}. Thus, the
recombinations from the stellar surface to the edge of the galactic halo
are not double-counted (Gnedin, private communication).

To summarize, our adopted standard model (SM) of reionization is
parametrized by the currently favored spatially flat cosmology described by
the parameter set: [$\sigma_8$, $n$, $h$, $\Omega_b$, $\Omega_\Lambda$,
$\Omega_{\rm M}$, $c_L$, $f_\star$, $f^{\rm H}_{\rm esc}$, $f^{\rm He}_{\rm
esc}$] = [0.9, 1.0, 0.7, 0.04, 0.7, 0.3, 30, 0.05, 0.05, 0.025], where $h$
is the Hubble constant in units of 100 km s$^{-1}$ Mpc$^{-1}$, and
$\Omega_b$, $\Omega_\Lambda$, and $\Omega_{\rm M}$ represent the
cosmological density parameters of baryons, the cosmological constant, and
matter respectively.  Our reionization model is described by the collapsed
baryon fraction, $F_B$, and the volume filling factors, $F_{\rm i}$, where:

\begin{equation}
F_B(z) = \rm{erfc}(\frac{\delta_c}{\sqrt 2 \sigma (R,z)})  
\end{equation}
\begin{equation}
F_{\rm i}(z) = \rho_B(z) \int_{25}^z dz_{on}
 \frac{dF_B}{dz}(z_{on}) \; \left[\frac {4 \pi}{3} r^3_{\rm{I,
 i}}(z_{on},z, M(z_{on}))\right].
\end{equation}

\noindent Here, $i$ corresponds to H~II or He~III, the critical overdensity
$\delta_c (z)$ equals 1.686 multiplied by a cosmology-dependent growth
factor \citep{ehu98}, $\rho_B(z)$ is the average IGM baryon density,
$z_{on}$ is the source turn-on redshift (we set the earliest redshift at
which star formation occurs to be 25), and $\sigma(R,z)$ is evaluated with
a spherical top-hat window function over a scale $R \propto M_{\rm h}$,
where $M_{\rm h}$ $\equiv$ $10^8 M_\odot [(1 + z_{on})/10]^{-1.5}$ is the
minimum total halo mass (the sum of DM and baryons) which has collapsed at
a source turn-on redshift $z_{on}$ utilizing H line cooling, and $M \equiv$
($\Omega_b/\Omega_{\rm M}$) $\times$ $M_{\rm h}$. We solve numerically for
the radius of the I-front, $r_{\rm I, i}$, whose evolution as a function of
$z_{on}$, $M$, $z$, and the time-dependent stellar ionizing fluxes is
described below.

We define reionization as the overlap of individual ionized regions of the
relevant species, i.e., when the volume filling factors of ionized hydrogen
and helium, $F_{\rm H II}$ = 1 and $F_{\rm He III}$ = 1. These are roughly
equivalent to the volume-averaged ionization fractions of each species if
almost all of the baryons are in the IGM up through reionization, which is
consistent with numerical simulations of the evolution of structure. We
emphasize that the semi-analytic treatment here defines reionization as the
overlap of fully ionized regions in the IGM, and corresponds to the
component of the IGM that dominates the ionization by volume filling factor
at high redshift.  By this definition, reionization precedes the GP
trough's disappearance, which represents the ionization of any remaining
H~I/He~II in overdense portions of the IGM or in individual ionized
regions.  Put another way, in the terminology of \citet{gnedin00}, the
model here is an accurate description of the ``pre-overlap'' and
``overlap'' phases of reionization, but not of the ``post-overlap'' epochs
which correspond to the ``outside-in'' ionization of neutral dense regions.
In a strict sense, the volume filling factors of H~II and He~III can never
equal unity exactly, owing to the presence of neutral regions in the
universe, whose evolution we do not track here, well after the reionization
of the IGM. Although we include the effects of IGM clumping in this paper,
the development of luminous objects and the gradual overlap of H~II regions
are themselves characterized only in an average sense. In reality, the
first astrophysical sources of ionizing photons are likely to be located in
dense regions embedded in the large-scale filamentary structure of matter,
so that reionization is a highly nonlinear, inhomogeneous process.  To
truly probe the complex details of this patchy reionization, one must turn
to numerical simulations, which can follow the detailed radiative transfer,
perform the necessary characterization of spatial variations, and reveal
the full 3D topology of reionization.

In Paper I, we presented newly calculated evolving spectra for Pop III
stars, which can then be converted into an ionizing photon rate as a direct
input to our reionization model. We reproduce this figure here from Paper I
for continuity. In Figure 1, we display the evolving spectra for continuous
and instantaneous star formation for synthetic clusters of Pop III stars,
as well as for a representative example of Pop II stars corresponding to
stellar metallicities, $Z = 0.001$ \citep{starburst}. The instantaneous
case converts $10^6$ $M_\odot$ into stars in a burst at time $t = 0$; in
the continuous case, gas is converted into stars at the steady rate of 1
$M_\odot$ yr$^{-1}$. The composite spectra (excluding nebular emission) are
shown at times of 1 and 15 Myr. We assume that the stars form in a Salpeter
IMF from 1--100 $M_\odot$, which is reasonable in the absence of a complete
theory of primordial star formation.

Figure 1 also shows the H~I and He~II ionizing photon production rates,
$Q_0$ and $Q_2$ respectively, for these synthetic clusters.  For the Pop~II
case, we plot the cluster $Q_2$ when Wolf-Rayet stars are included and
excluded. This provides an indication of the role played by stellar mass
loss for the gain in He~II ionization from Pop~II to Pop~III. This gain
could be lowered if there are significant numbers of Wolf-Rayet stars at
Pop~II metallicities. As we discussed in Paper I, the presence of the
Wolf-Rayet phenomenon at high redshift and low metallicity is thought to be
unlikely. We therefore focus in this paper on directly comparing the
ionizing radiation from the main-sequence phases of Pop~II and Pop~III;
this will roughly bracket the range of their effects on reionization,
particularly for He~II.  From Figure 1, the Pop III cluster has 60\%
stronger H~I ionization and $10^5$ times more He~II ionizing photons
relative to the Pop II cluster with the same IMF and total mass, and
excluding Wolf-Rayet stars. These differences, particularly for He~II,
could have potentially large effects for the reionization of the IGM, which
we examine in the next section. Also displayed in the upper right panel of
Figure 1 are the instantaneous values of $Q_2$ from the zero-age main
sequences corresponding to stellar carbon abundances, $Z_C = $ 10$^{-8}$,
10$^{-7}$, and 10$^{-6}$, marked with filled squares.  These are intended
to mimic the effects of trace levels of $^{12}$C in the stellar core, and
to represent a possible ``second generation'' of stars. These points show
the sharp decline in the He~II ionizing photon production when small
abundances of $^{12}$C are available to stars.

We provide here the fits\footnote{We refer the reader to \citet{schaerer}
for a detailed treatment and fits of the time-averaged ionizing photon
rates as a function of stellar mass for $Z = 0$ stars. The IMF-averaged
ionizing fluxes provided there as a function of time are in agreement with
our results here.} for the logarithm of the ionizing photon rates,
log($Q_0$) and log($Q_2$), of a $10^6$ $M_\odot$ Pop III cluster in the
instantaneous burst case (see \S 3 for the continuous star formation case);
these are good to within a few percent over the fitted range.  The time
variable is $T \equiv \log t$, where $t$ is the time in years since the
cluster turned on. For $t = 0 - 2.8$ Myr,
\begin{equation}
\log(Q_0) = 54.16 - 0.5T + 0.053T^2.
\end{equation}
For $t = 2.8 - 30$ Myr,
\begin{eqnarray}
\log(Q_0) &=& -128.24 + 175.1T - 17.92(3T^2 - 1) + 1.35(5T^3 - 3T)
\nonumber
\\ & &- 0.0086(35T^4 - 30T^2 + 3) - 0.21 \times \exp[-0.5(T -
7.22)^2].
\end{eqnarray}
After this time, $Q_0$ drops to about 1\% of its value at $t = 0$, and we
set $Q_0 = 0$ for convenience, although this is not strictly true. For
He~II, from $t = 0 - 0.9$ Myr,
\begin{equation}
\log(Q_2) = 50.72 + 0.235T - 0.027T^2,
\end{equation} for $t = 0.9 - 2.55$ Myr,
\begin{equation}
\log(Q_2) = -{\rm exp}[31.33(T - 6.44)] \times (-0.047T + 5.13) -
(T - 6.08)^2/0.43 + 0.77,
\end{equation} and we set $Q_2 = 0$ after 2.55 Myr.

The growth of individual ionized regions is a function of the
time-dependent source luminosity and the stellar cluster's turn-on
redshift, and their evolution is given by the balance between
photoionization and recombination in an expanding IGM \citep{sg87,
donshull87}. The differential equations describing the evolution of the
I-fronts' radii for H~II and He~III, $r_{\rm{I},H} (t)$ and $r_{\rm{I},He}
(t)$, in equation (2) are:

\begin{equation}
n_{\rm H}(t) \left[\frac{ d r_{\rm{I},{\rm H}}}{dt} - H(t) r_{\rm{I},{\rm H}}
\right] = \frac{1}{4 \pi
  r_{\rm{I},{\rm H}}^2} \, \left[f_\star f^{\rm H}_{\rm esc} Q_0(t) - \frac{4
\pi}{3}
  \alpha_{\rm HI}^{\rm B} c_L n_e(t)  n_{\rm HII}(t) r_{\rm{I},{\rm H}}^3 \right]
\end{equation}
\begin{equation}
n_{\rm He}(t) \left[\frac{ d r_{\rm{I},{\rm He}}}{dt} - H(t) r_{\rm{I},{\rm He}}
\right] = \frac{1}{4
  \pi  r_{\rm{I},{\rm He}}^2} \, \left[f_\star f^{\rm He}_{\rm esc} Q_2(t) -
\frac{4 \pi}{3}
  \alpha_{\rm HeII}^{\rm B} c_L n_e(t)  n_{\rm HeIII}(t) r_{\rm{I},{\rm He}}^3.
\right]
\end{equation}

We assume case B recombination for both H~II and He~III in the
pre-reionization IGM, since the Ly$\alpha$ and recombination line photons
for both of these are likely to be resonantly scattered and absorbed
locally.  The line and continuum recombination photons to the $n = 1$
states, if sufficiently redshifted, could potentially be relevant for the
global radiation field at subsequent epochs, especially for case B
recombination. This is unlikely, however, to be a large effect prior to the
full reionization of the IGM. A realistic description of recombination will
lie somewhere between case A and case B. We proceed here with the
assumption of case B recombination, and reduce the above equations to:

\begin{equation}
\frac{d r^3_{\rm{I},{\rm H}}(t)}{dt} = 3 H(t) r^3_{\rm{I},{\rm H}}(t) +
\frac{3}{4 \pi
  n_{\rm H}(t)} \left[f_\star f^{\rm H}_{\rm esc}  Q_0(t) - \frac{4 \pi}{3}
  r_{\rm{I},{\rm H}}^3 \alpha_{\rm HI}^{\rm B} c_L n_e(t) n_{\rm H II} (t) \right]
\end{equation}
\begin{equation}
\frac{d r^3_{\rm{I},{\rm He}}(t)}{dt} = 3 H(t) r^3_{\rm{I},{\rm He}}(t) +
\frac{3}{4 \pi
  n_{\rm He}(t)} \left[f_\star f^{\rm He}_{\rm esc} Q_2(t) - \frac{4 \pi}{3}
  r_{\rm{I},{\rm He}}^3 \alpha_{\rm HeII}^{\rm B} c_L n_e(t) n_{\rm He III} (t)
\right].
\end{equation}

At each redshift, we solve numerically for the growth of the I-fronts with
a fourth-order Runge-Kutta method for all the preceding source turn-on
redshifts, each corresponding to $t = 0$ in equations (9) and (10). The
time steps at each redshift are set to be 5--10\% of the recombination
timescale appropriate for H~II or He~III at that redshift. These were the
largest time steps that ensured numerical convergence of the results.  The
recombination timescale is given by:

\begin{equation}
\bar{t}_{\rm rec} (z) = \frac{1}{\alpha^{\rm B} c_L (1 + 2x_{\rm He}) \bar{n}_{\rm H} (z)},
\end{equation}

\noindent where $\bar{n}_{\rm H} (z)$ is the average total number density
of hydrogen in the IGM ($\bar{n}_{\rm H}$ ($z=0$) $\sim$ 1.7 $\times
10^{-7}$ cm$^{-3}$ in our SM), $x_{\rm He}$ is the ratio of He to H by
number, about 0.0789 for $Y_{\rm He}$ = 0.24, $n_e$ = $n_{\rm HII}$ +
2$n_{\rm He III}$ = $(1 + 2x_{\rm He})$ $n_{\rm H}$, under the assumption
that all H (He) in the vicinity of the I-front is in the form of H~II
(He~III), and $\alpha^{\rm B}$ is the case B recombination coefficient for
H~II or He~III. By virtue of the definition of $c_L$, equation (11) assumes
a spatially constant $\alpha^{\rm B}$. Furthermore, since we do not solve
in detail for the thermal evolution of the IGM in the vicinity of the
I-fronts, we assume a temperature of $10^4$ K and take $\alpha_{\rm
HI}^{\rm B}$ ($10^4$ K) = 2.59 $\times$ $10^{-13}$ cm$^3$ s$^{-1}$, with
$\alpha_{\rm HeII}^{\rm B}$ ($10^4$ K) = 5.83 $\alpha_{\rm HI}^{\rm B}$
($10^4$ K) \citep{spitzer}.

This method of solving for the nonequilibrium growth of the I-fronts of
H~II and He~III is a significant improvement on the treatment in V2000,
which used the analytic solutions for $r_{\rm I, i} (t)$ provided by
\citet{sg87}. The drawback of that method was the underlying assumption of
a constant source luminosity. In the present treatment, the evolution of
the I-fronts is followed more accurately, and we can distinguish between
the cases of continuous and bursty star formation.

\section{Results}

In this section, we focus on the effects of the first stars for
cosmological reionization, using the model described in \S 2. Specifically,
we consider combinations of evolving Pop III and Pop II
spectra, in continuous and bursty star formation scenarios, whose
definitions directly correspond to those in Figure 1. In the case of
continuous star formation, we take the appropriate values of $Q_0$ and
$Q_2$ for Pop III or II at a time of 10$^6$ yr from Figure 1, when
the cluster stellar masses equal $10^6$ $M_\odot$. We do this to be
consistent with the case of synthetic spectra for a burst of star formation
for a 10$^6$ $M_\odot$ cluster: the star formation rate for the continuous
case being 1 $M_\odot$ yr$^{-1}$, the total gas mass converted to stars
equals that for bursty star formation at a time of 1 Myr. For Pop
III, $Q_0 = 1.04 \times 10^{53}$ s$^{-1}$ and $Q_2 = 1.56 \times 10^{51}$
s$^{-1}$. For the adopted Pop II cluster, $Q_0 = 7.76 \times
10^{52}$ s$^{-1}$ and $Q_2 = 1.32 \times 10^{46}$ s$^{-1}$.

The cases considered below are intended to provide an indicative range for
the effects of Pop III stars on the reionization of the high-$z$ IGM, given
reasonable values for the many parameters that describe the model. As
emphasized earlier, the issues of the sources of reionization, and the
formation sites and cosmological significance of the first (presumably
metal-free) stars are at present unresolved. Furthermore, the nature of
high-$z$ star formation is completely unknown, and may not separate cleanly
into bursty versus continuous modes. Thus, the distinction drawn below
between Pop III and Pop II is convenient rather than exact, owing to the
lack of detailed evolving spectra in the literature for stellar
metallicities between $\sim$ 0.001 and 0 (\citealt{schaerer} has computed
the case of $Z = 0.0004$ stars, albeit with mass loss and for a different
range of the stellar IMF than we consider here). We present the results
below with the understanding that reality lies somewhere between the
considered cases.

\subsection{Pure Pop III stellar spectra}

We begin by considering the extreme case in which Pop III stellar spectra
are assumed for all star formation through reionization. This is clearly
not realistic; the reader may take the results here to represent the most
extreme effects of metal-free stars. In Figure 2, we display the redshift
evolution of $F_{\rm B}$, $F_{\rm HII}$, and $F_{\rm HeIII}$ for the SM
with continuous and bursty star formation with evolving Pop III spectra. We
do not display the evolution of $F_{\rm HII}$ and $F_{\rm HeIII}$ here or
in subsequent figures after these quantities equal unity, since the model
in this work describes reionization accurately only up through the overlap
stage. The values of these three quantities are also shown in Table 1 at a
series of sample redshifts, for the various cases depicted in Figure 2.

For continuous (bursty) Pop III star formation, H~I
reionization\footnote{Although we do not include low-mass halos in this
work for the reasons detailed in \S 2, the details of the radiative
feedback on early starforming halos are not yet quantitatively resolved. As
a comparison with the results in this section, we find that the general
effect of including halos with virial temperature above $10^3$ K (rather
than $10^4$ K) is an increased reionization redshift. Specifically, for
continuous (bursty) Pop III star formation, H~I reionization occurs at $z
\sim 11.8$ (6.2), and that of He~II at $z \sim 6.3$ (0.7).} occurs at $z
\sim 9$ (4.7), and that of He~II at $z \sim 5.1$ (0.7). As may be expected,
reionization occurs later for both H~I and He~II in the case of evolving
spectra from an instantaneous burst relative to those from the continuous
case. This can be attributed to the decline in ionizing photons at times
exceeding a few million years as the burst fades. A critical question,
which we have sidestepped here, is what determines the time lag between
bursts of star formation at early epochs. Clearly, in the limit of zero
lag, there will be no distinction between continuous and bursty modes, as
far as the emergent spectra for the IGM are concerned. Since this issue is
unresolved, Figure 2 shows the range of the effects of Pop III stars alone
on reionization for both star formation modes. For both H~I and He~II, this
range contains the current observational values of their respective
reionization epochs, $z_{\rm reion, H}$ $\sim 6$ and $z_{\rm reion, He}$
$\sim 3$.

\subsection{Pop III switching to Pop II}

A more realistic description of the role played by stars in reionization
should involve stellar populations of non-zero metallicity in addition to
Pop III. This translates into solving the problem of the complex interplay
between cosmological metal transport and the mixing timescales in the ISM
of individual high-$z$ halos, in which the metals generated by the first
population of stars are incorporated into subsequent stellar
generations. This is, however, likely to be influenced heavily by the local
balance between stellar/SN feedback and the availability of coolants, a
problem whose inherent dependence on spatial inhomogeneity places it beyond
the scope of this work. We attempt below to evaluate the effect of Pop III
($Z = 0$) stars, which then switch to Pop II ($Z = 0.001)$ stars at some
later time, with the caveat that this definition of Pop II is one based on
the available spectral templates of non-zero metallicity. We stress that
the transition from Pop III to Pop II is driven by the physical condition
of a metallicity threshold rather than a fixed timescale for halo
self-enrichment.  In reality, Pop II star formation may span a range of
metallicities between $Z = 0$ and $Z = 0.001$, corresponding to many
generations of stars. The transition from Pop III to Pop II may occur at,
e.g., a metallicity of $10^{-4}$ $Z_\odot$ as some authors have found
(\citealt{schneider}, and references therein). For our purposes below, the
important distinction is between stellar populations that generate He~II
ionizing photons, and those that do not. This rough division is justified
by the steep dependence of $Q_2$ on the stellar metallicity (Figure 1).

An estimate of the time at which Pop III star formation ceases and that of
Pop II begins involves assumptions about many physical processes which are
not yet well understood, such as the mixing and re-incorporation timescales
for the generated metals.  We consider only continuous star formation,
because for bursts it becomes difficult to define an average global time
when Pop II star formation begins. Moreover, pockets of metal-free gas
cannot be ruled out at any epoch.  In addition, the nature of the
Press-Schechter method complicates the treatment of bursty star
formation. The formalism is designed to track which mass overdensities are
going nonlinear at any redshift, but it does not have information on the
detailed stellar history of the baryons in halos at those epochs. Thus, for
an individual halo, the connection between the time elapsed from the onset
of Pop III star formation and the age of the universe is made more
consistently for continuous rather than bursting modes. This would
correspond to a ``universal'' self-enrichment timescale. This timescale
could be increased by the preferential ejection of metals into the IGM
through SN feedback or decreased if the metals cool rapidly and exist
predominantly in a cold ISM phase rather than a hot phase. Given these
uncertainties associated with the incorporation of freshly synthesized
metals into new stars, we will consider two ISM mixing timescales that span
an order of magnitude, $10^8$ and $10^9$ yr \citep{maclow}.

In Figure 3, we display the redshift evolution of $F_{\rm HII}$ and $F_{\rm
HeIII}$ for the cases of Pop III stars lasting for $10^8$ and $10^9$ yr,
after which the ionizing spectra switch to Pop II. As in \S 3.1, the values
of these quantities are shown in Table 1 at a series of redshifts for the
SM cases depicted in Figure 3. In addition, we show a case for $F_{\rm
HeIII}$ where Pop III spectra switch to Pop II at $10^9$ yr, with $c_L$ =
30 for times up through $10^9$ yr, and $c_L$ = 1 at times exceeding $10^9$
yr. We do this in order to provide an indication of the fate of those IGM
regions which are not strongly clumped, after they have experienced partial
ionization from early Pop III stellar activity. Such regions, with
overdensities of order 1 to a few and $c_L$ $\sim$ 1, are likely to be the
dominant component of the IGM by volume at late epochs. Their evolution,
once the Pop III ionizing sources turn off, is dominated by recombinations
and is hence sensitive to the chosen value of $c_L$.

We find that H~I reionization occurs at $z \sim 9.0$ (8.7) for Pop III star
formation lasting for $10^9$ ($10^8$) yr. This small difference in the
reionization epochs is not surprising, given that the H~I ionizing photon
rates between Pop III and Pop II differ only by about 60\%.  In addition,
since the age of the universe exceeds $10^9$ yr only at $z \la 6$, the case
of Pop III stars lasting for $10^9$ yr is effectively no different from
that of considering only Pop III spectra (Figure 2, continuous star
formation case).

The duration of Pop III activity has, however, more dramatic consequences
for He~II reionization, given the many orders of magnitude difference
between the Pop III and Pop II He~II ionizing fluxes. Although He~II never
reionizes completely in any of the cases, the scenario with a $10^9$ yr Pop
III timescale reaches ionization fractions of about 60\% by $z \sim 5.6$
when Pop III stars are turned off. Subsequent to that, the ionization of
He~III plummets to very low values for $c_L$ = 30, effectively to the Pop
II curve, and recovers mildly by $ z \sim 0$. Although the redshift of the
precipitous drop in $F_{\rm HeIII}$ is an artifact of the Pop III timescale
considered here, the strong decrease in $F_{\rm HeIII}$ itself is real,
since it is tracked over about a hundred time steps over $\Delta z \sim$
0.5. For the case with a $10^9$ yr Pop III timescale where $c_L$ = 30 (1)
up through (after) the He~III ionization peak at $z \sim 5.6$,
recombinations are less effective at $z \la 5.6$. Hence, the He~III
ionization fraction declines more gradually and remains at levels of about
3\% at $z \sim 0$. Together, these two cases provide an indicative range of
the evolution of IGM regions which have experienced partial He~II
ionization by Pop III stars in our reionization model.

From the results presented through this point, we see that He~II may
experience full or partial ionization followed by recombination, depending
on the duration of Pop III stellar activity, prior to its currently
detected reionization epoch of $z \sim 3$.  The fate of any ionized relic
He~III regions, particularly in the IGM voids, is discussed further in \S
3.4.

\subsection{The Amount of Small-Scale Power}

The redshift, $z_{\rm reion}$, at which H~I or He~II reionization occurs is
sensitive to the amount of small-scale power available in the structure
formation model for the first luminous objects. This epoch is determined
primarily by two cosmological parameters, $\sigma_8$ and $n$, which we use
as inputs to quantify reionization. An increase in either of these
parameters directly leads to enhanced small-scale power and earlier
reionization; in particular, $z_{\rm reion}$ is affected strongly by $n$
(see, e.g., \citealt{venk02}) for the case of H~I reionization. Thus, even
if we assume that all the other parameters characterizing reionization are
well-constrained, $n$ and $\sigma_8$ have to combine in the right way to
meet the observational limits on $z_{\rm reion}$. In this era of precision
cosmology, most cosmological parameters are becoming highly constrained,
with $\sigma_8$ remaining as one of the few parameters with significant
uncertainty. As noted earlier, the data on $\sigma_8$ appear to be divided
into two values, a high value of about 0.9 -- 1.0
\citep{evrard,pen,refregier}, and a low value of about 0.7 \citep{reip,
seljak}, each with a claimed error bar of about 10\%. In this section, we
examine the consequences of a low value for $\sigma_8$ for $z_{\rm reion,
H}$ (we chose $\sigma_8 = 0.9$ for our SM), given the current uncertainty
in the value of $n$. We focus on H~I in particular, as the detection of
$z_{\rm reion, H}$ is the next observational frontier of interest for
reionization. This quantity can potentially be a useful probe of the
small-scale power in a reionization model. We emphasize that we are not
assuming that the astrophysics of reionization is completely
understood. Rather, we are interested in the consequences of lowering the
value of $\sigma_8$ for reionization, given reasonable values for the
astrophysical parameters in our reionization model.

The current limits on $n$ are about 0.9 -- 1.1 from a combination of data
on the CMB, large scale structure, and the Ly$\alpha$ forest (see, e.g.,
\citealt{wangtz01, dasi}), with a preference in this range for $n$ $\sim$
0.9 -- 1.0 from analyses of joint data sets. In Table 2, we show the values
of $z_{\rm reion, H}$ for $\sigma_8$ = 0.7 and $n$ = 0.9, 1.0, and 1.1, for
the two cases of continuous and bursty Pop III star formation. The
corresponding numbers for $\sigma_8$ = 0.9 are shown in parentheses for
each case. As emphasized earlier, the case of continuous star formation is
intended to demonstrate the most extreme effects of $Z = 0$ stars. The
bursty case provides a lower limit to $z_{\rm reion, H}$ here, but its
inherent dependence on the unknown time lag between bursts implies that
even lower values of $z_{\rm reion, H}$ are possible.

When we compare the values of $z_{\rm reion, H}$ from Table 2 with current
data, the following cases appear to be inconsistent with observations,
within the uncertainties of our reionization model: $0.9 \leq n \leq 1.1$
with $\sigma_8$ = 0.7 or $\sigma_8$ = 0.9 (bursty case); $n = 0.9$ with
$\sigma_8$ = 0.7 (continuous case); and, $n = 1.1$ with $\sigma_8$ = 0.9
(continuous case).  Those of the excluded cases with $n$ = 0.9 are worth
noting, since this value for $n$ is preferred by the most recent joint data
set analyses such as \citet{wangtz01} and \citet{tzh01}.  Note that we
treat reionization through the overlap phase, but not the ``clearing out''
of the last dense neutral regions in the IGM. Thus, the numbers in Table 2
are, in this sense, an upper bound to the redshift of reionization as
determined by GP troughs. A more detailed treatment using numerical
simulations would result in a slightly downward revision of $z_{\rm reion,
H}$ in Table 2.

We emphasize that the relation between $n$, $\sigma_8$ and $z_{\rm reion,
H}$ as shown in Table 2 is oversimplified, given the uncertainties in our
reionization model. These include the unknown parameters quantifying Pop
III star formation (such as the IMF, mode of star formation, and the
maximum stellar masses in the IMF). Additionally, the matter power spectrum
may not be described on all scales by a single value of $n$, and the
feedback from stars and SNe on baryons in halos could partially erase the
correlation between small-scale power and $z_{\rm reion, H}$.
Alternatively, $z_{\rm reion, H}$ can be raised by increasing $f_\star$ or
$f^{\rm H}_{\rm esc}$ in the model. This would require star formation to be
more efficient and/or more widespread than we have assumed in the SM, or
require the escape fraction of ionizing radiation to be larger. This is
something that observations of the local universe do not support, although
we cannot rule this out. Our main point here is that the relation between
$n$, $\sigma_8$, and $z_{\rm reion, H}$, subject to the above caveats, can
be useful if combined with future observational determinations of $z_{\rm
reion, H}$ and even slightly tighter limits on $n$. In particular, this
technique potentially has the power to rule out the combination of low
values of $\sigma_8$ $\la$ 0.7 and $n < 1$.

\subsection{The Fate of Primordial He~III Regions}

The possibility of an early epoch of partial He~II reionization in the IGM
by Pop III stars begs the question of what happens to these regions once
the Pop III epoch ends. If we can predict the evolution of these early
He~III regions, we can compare them with the best available constraints on
He~II ionization in the high-redshift IGM \citep{kriss,shull02}. If the
He~II in the IGM is reionized shortly after H~I reionization is
accomplished, will these He~III regions recombine by $z \sim 3$, where we
have observational constraints?

In a clumpy IGM, we can write the recombination time for He~III as
a function of the local overdensity $\delta$, where $\delta =
n_{\rm H}/\langle n_{\rm H} \rangle$, and $\langle n_{\rm H}
\rangle$ is the average total hydrogen density in the IGM. We take
a critical mass density $\rho_{crit} = 1.878 \times 10^{-29}h^2$ g
cm$^{-3}$ and a primordial He mass fraction $Y=0.24$ to calculate
$\langle n_{\rm H} \rangle$. Thus,
\begin{equation}
n_{\rm H}(z) = (1.71 \times 10^{-7} {\rm cm}^{-3})(1+z)^3 \left(
\frac{\Omega_b h^2}{0.02} \right) \delta .
\end{equation}
The recombination time, $t_{\rm rec}$, of He~III to He~II is given by
$t_{\rm rec} = (n_e \alpha_{\rm HeII}^{B})^{-1}$, where the Case B
recombination coefficient for He~II can be written with an explicit
temperature dependence, $\alpha_{\rm HeII}^{B} = 1.51 \times 10^{-12}
(T/10^4\, {\rm K})^{-0.7}$ cm$^{3}$ s$^{-1}$.  Combining these
relationships and taking $n_e = 1.16n_{\rm H}$ for fully ionized gas with
He/H = 0.0789 by number, we derive the local recombination time as a
function of redshift and overdensity\footnote{\citet{go97} and
\citet{mad99} define a spatially-averaged recombination time in terms of a
clumping factor $C_{\rm HII} = \langle n^2_{\rm HII}\rangle / \langle
n_{\rm HII}\rangle^2$. Their formalism does not follow the recombination of
the IGM at a point with a single overdensity. We are interested here in
distinguishing between the ``filaments'' with high $\delta$ and ``voids''
with low $\delta$ in observations where the local overdensity is mapped by
the Ly$\alpha$ forest. Thus, we define the recombination time here in terms
of the local overdensity rather than a spatially-averaged clumping
factor.},
\begin{equation}
t_{\rm rec} = (1.31 \times 10^{8}\, {\rm yr}) \left(
\frac{1+z}{10} \right)^{-3} \left( \frac{\Omega_b h^2}{0.02}
\right)^{-1} \left( \frac{T}{10^4\, {\rm K}} \right)^{0.7}
\delta^{-1}.
\end{equation}
In Figure~\ref{fig:recfig} we plot the values of four times this
expression, and compare it to the time from the reionization of He~II to $z
= 3$, where we have observational constraints on the ionization of He in
the IGM. We choose to display 4$t_{\rm rec}$ in order to indicate a
recombination e-folding timescale over which the evolving IGM ionization
fraction changes appreciably ($e^{-4} \sim$ 0.02, whereas $e^{-1} \sim$
0.37). We have explored the possibility that the He~II in the IGM was first
reionized to He~III at $z \sim 5$ by zero-metallicity stars.  Figure 4
shows that the ``void" regions with $\delta \la 0.3$ would not have time to
recombine before $z = 3$.  Such ``relic'' He~III regions may be visible to
sensitive observations of the He~II GP effect.  However, the large numbers
of quasars being discovered at redshifts $z \geq 4$ \citep{fan3} are likely
to provide sufficient ionizing radiation to affect the He~II ionization
fraction, $f_{\rm HeII} = n({\rm He~II})/n({\rm He})$.  Because $f_{\rm
HeII} \ll 1$, a small fraction of He~III recombinations, together with
photoionization by QSO radiation, can shift the He~II ionization and the
He~II/H~I ratio, $\eta$.  The detailed ionization history would require a
careful integration of the non-equilibrium H~I and He~II chemistry, which
is beyond the scope of this paper.

\citet{fardal} (see also \citealt{mirost92}) define the He~II to H~I column
density ratio to be:
\begin{equation}
\eta \equiv \frac{n_{\rm HeIII}}{n_{\rm HII}} \frac{\alpha_{\rm
HeII}}{\alpha_{\rm HI}} \frac{\Gamma_{\rm HI}}{\Gamma_{\rm
HeII}}.
\end{equation}
This ratio is sensitive to the shape of the ionizing spectrum. Paper I
derived the intrinsic $\eta$ of metal-free stars and the low-metallicity
ZAMS.  We find that $\eta = 10$ for the Pop III ZAMS cluster and remains
below 50 for 2.5 Myr. By comparison, $\eta = 20$ for a composite QSO with
power-law spectral index $\alpha = 1.8$ \citep{telfer}, and $\eta = 10$ for
$\alpha = 1.3$.  The low $\eta$ for the Pop III cluster results from its
unusual spectrum. The H~I photoionization rate $\Gamma_{\rm HI}$ is
proportional to the integral above 1 Ryd of the specific photon flux times
the ionization cross section, [$F_{\nu} \sigma_{\nu}/ h \nu$]. For the
power-law spectrum, $F_{\nu}$ attains a maximum at 1 Ryd, where the cross
section peaks. The Pop III composite spectrum peaks near 3 - 4 Ryd, where
the H cross section is $\sim$64 times smaller (for $\sigma_\nu \propto
\nu^{-3}$). We obtain a lower H~I photoionization rate for the Pop III
spectrum and, when taken together with its strong He~II ionization, a
correspondingly lower $\eta$.

In a study of the He~II GP effect toward HE 2347-4342 with FUSE at $z = 2.3
- 2.9$, \citet{kriss} found several regions of unusually low $\eta$, a
signature of hard ionizing sources. Many of these regions can be explained
by the contemporaneous presence of QSOs with hard power-law spectra
\citep{telfer, kriss}. In a more detailed analysis of the FUSE data,
\citet{shull02} also find that in ``void'' regions with $\log N({\rm H~I})
< 12.3$, $\eta$ is systematically higher than 100.  These regions can be
explained by the presence of more broadly distributed soft sources, such as
dwarf galaxies, although this does not explain the large fluctuations in
$\eta (z)$. In this picture, the hard sources responsible for low $\eta$
(QSOs) lie primarily in regions of high neutral hydrogen abundance.

The early He~II reionization scenario provides an alternative explanation
for the high $\eta$ regions. If He is ionized to He~III at $z \sim$ 6,
regions with $\delta > 1.0$ will recombine by $z = 3$, erasing all
signature of the first ionization. However, in the underdense regions with
$\delta < 1.0$, there will be time for only partial recombination of He~III
to He~II. These ``relic'' He~III regions may be detected by sensitive
observations of the He~II GP effect at high redshift. We note, however,
that the rise in the large bright QSO population at $z \la 6$, which we
have not accounted for here, will influence the ionization equilibrium of
He~II and H~I in such relic regions. It would then become more challenging
to detect this potential signature of metal-free stellar activity. Tests of
this idea must await efforts with the {\it Cosmic Origins Spectrograph} on
the {\it Hubble Space Telescope} to push observations of the He~II GP
effect to $z > 3$.

\section{Conclusions}

We showed in Paper I that Pop III stars have unusually hard spectra
and elevated H~I and He~II ionizing photon rates. These properties
motivated this work, where we examined the role played by these objects in
H~I and He~II reionization through a semi-analytic reionization model
described by a reasonable set of parameters in the currently favored
cosmology. Our general conclusions are:

\begin{itemize}

\item[1.] We find that Pop III stars can be cosmologically significant for
reionization, particularly for He~II. For Pop III alone, H~I and He~II
reionize at redshifts $z_{\rm reion, H}$ $\sim$ 9.0 (4.7) and $z_{\rm
reion, He}$ $\sim$ 5.1 (0.7) for continuous (bursty) modes of star
formation.

\item[2.] We also considered a more realistic scenario involving a Pop III
phase of (continuous) star formation which switches to Pop II after a
self-enrichment timescale for primordial star-forming gas. We find that H~I
reionization occurs at $z_{\rm reion, H}$ $\sim$ 8.7 or 9.0, depending on
whether the Pop III stage lasts $10^8$ or $10^9$ yr respectively. He~II
never reionizes completely in either case, although the ionization fraction
of He~III reaches a maximum of about 60\% at $z \sim 5.6$ for a $10^9$ yr
self-enrichment timescale.

\item[3.] Since the reionization epoch is sensitive to the power available
on small scales, data on H~I reionization can critically test, and possibly
rule out, low values of $\sigma_8$ ($\la$ 0.7), particularly if $n
<$ 1.

\end{itemize}

By measuring $z_{\rm reion, H}$ from the CMB and high-$z$ spectroscopic
studies, and by using direct imaging techniques to detect Pop III stellar
clusters (Paper I), one can strongly constrain the role played by Pop III
stars in H~I reionization. The current evidence for a complete H~I GP
trough, and hence $z_{\rm reion, H}$, comes from the spectrum of a single
$z = 6.28$ QSO.  The acquisition of more GP data along more lines of sight
to sources at $z \sim$ 6 -- 9 is required to adequately represent how the
appearance and duration of the GP trough varies with redshift.  Such
observations are within the capabilities of the {\it Sloan Digital Sky
Survey}, which should detect about 20 bright quasars at $z \ga$ 6 during
the course of the survey \citep{becker}, and are important targets in the
planning of the {\it Next Generation Space Telescope}.

The significance of Pop III stars for He~II reionization can be tested by
future measurements of the He~II GP effect in the IGM at $z \sim$ 3 -- 5,
particularly in underdense regions of the IGM, which may not have had the
time to recombine by $z \sim 3$ after experiencing ionization by Pop III
stars at early times. These relic ionized voids may retain the unique
spectral imprint of the first, metal-free stellar populations.

\acknowledgements

We thank our referee, Andrea Ferrara, and Mark Giroux for helpful
suggestions which improved the manuscript.  We gratefully acknowledge
support from NASA LTSA grant NAG5-7262 and FUSE contract NAS5-32985 at the
University of Colorado.

\clearpage

\begin{figure}
\plotone{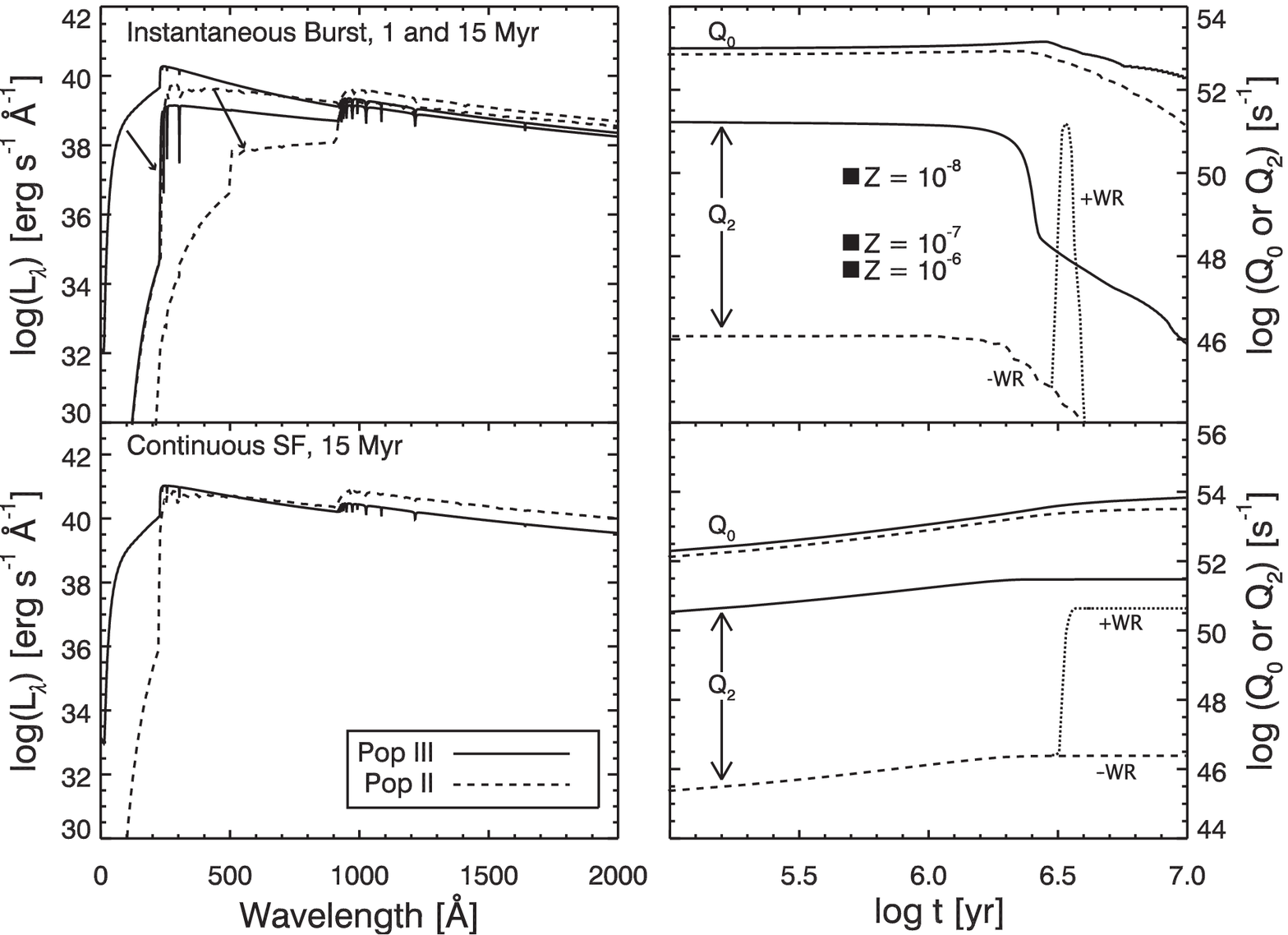}
\caption{Comparison of evolving spectra, $Q_0$ and $Q_2$, for synthetic
Pop~II and Pop~III stellar clusters. Upper left: Composite spectra for
Pop~II (dashed) and Pop~III (solid) clusters at times of 1 and 15 Myr after
converting $10^6$ $M_\odot$ into stars in a Salpeter IMF from 1--100
$M_\odot$ in an instantaneous burst. At 15 Myr the Pop~II spectrum has
faded in H~I ionization, but the Pop~III cluster is still a significant
source of H~I ionization, owing to the presence of $Z = 0$ stars with $M =
10 - 15$ \Msun\ and $T_{\rm eff} > 50,000$ K. No nebular emission is
included here. Lower left: example spectra for the continuous star
formation case at 15 Myr, which forms stars in a Salpeter IMF from 1--100
$M_\odot$ at the rate of 1 $M_\odot$ yr$^{-1}$. Upper and lower right:
IMF-weighted cluster $Q_0$ and $Q_2$ corresponding to the
instantaneous/continuous cases at left. For Pop~II, the cluster $Q_2$ is
plotted for the cases when Wolf-Rayet stars are included (dotted) and
excluded (dashed), indicating the effects of stellar mass loss on the gain
in He~II ionization from Pop~II to Pop~III.  In the upper right panel we
mark with filled squares the instantaneous values of $Q_2$ from the
zero-age main sequences with carbon abundances, $Z_C = $ 10$^{-8}$,
10$^{-7}$, and 10$^{-6}$. These points show the sharp decline in He~II
ionizing photon production when small abundances of $^{12}$C are included.}
\label{fig:specfig}
\end{figure}

\clearpage

\begin{figure}
\includegraphics[height=3.5in]{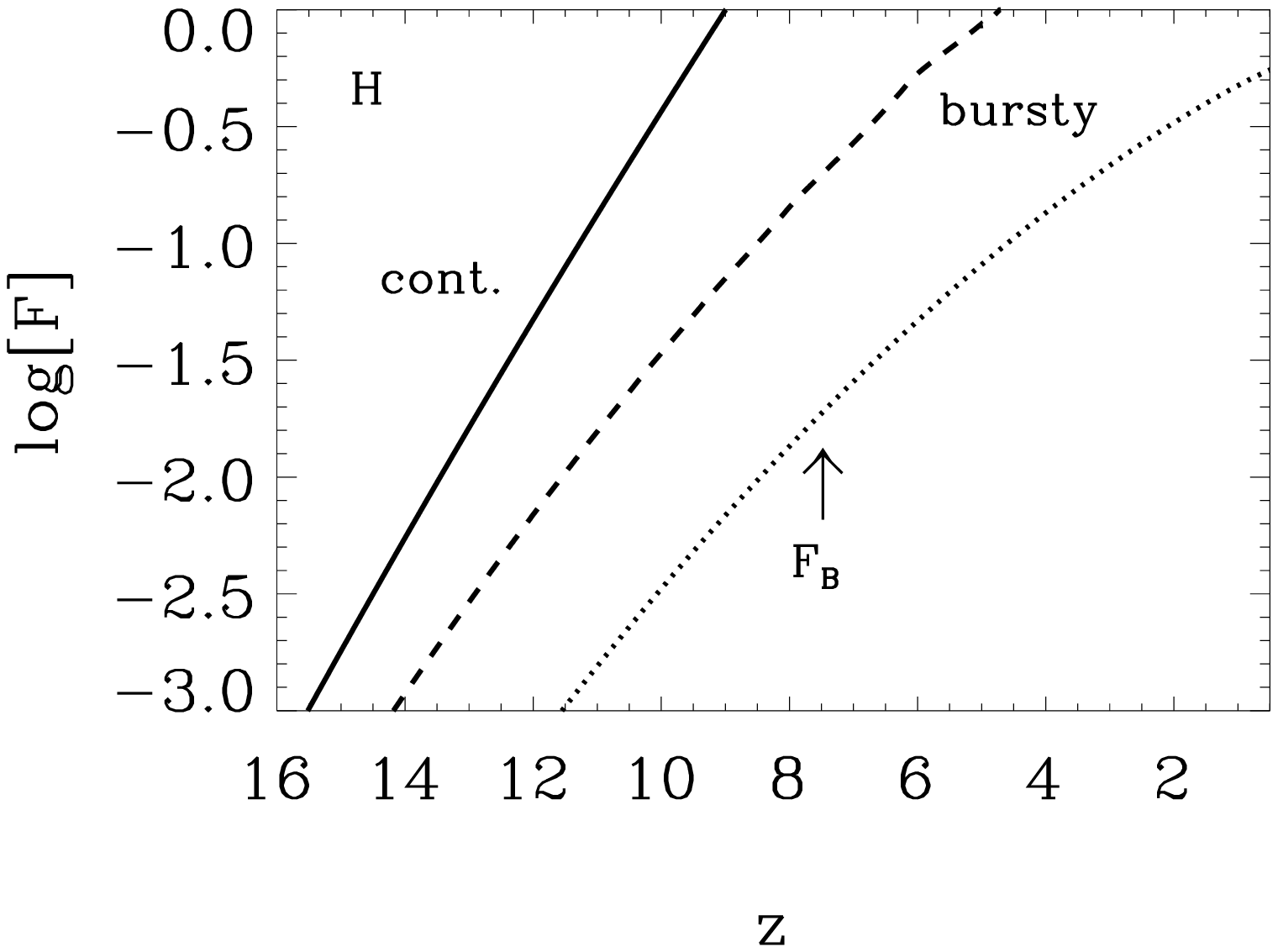}
\includegraphics[height=3.5in]{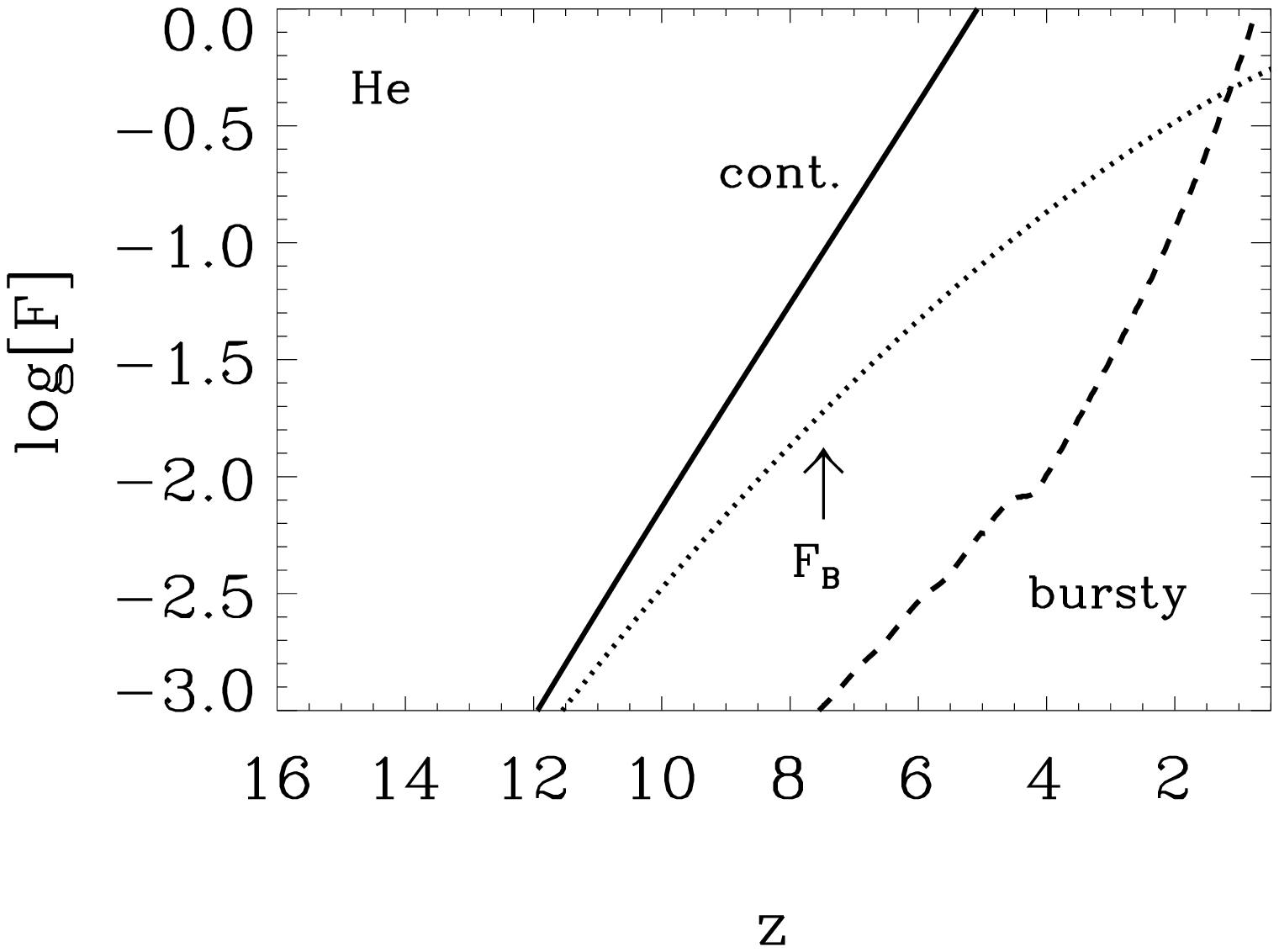}
\caption{The redshift evolution of the fraction of baryons in star-forming
halos $F_{\rm B}$ (dotted lines), and the volume filling factors of H~II
and He~III, $F_{\rm HII}$ and $F_{\rm HeIII}$, for the standard model in
this work: $\sigma_8$ = 0.9, $\Omega_b$ = 0.04, $h$ = 0.7, $n$ = 1.0,
$\Omega_\Lambda$ = 0.7, $\Omega_{\rm M}$ = 0.3, $c_L$ = 30, $f_\star$ =
0.05, $f^{\rm H}_{\rm esc}$ = 0.05, $f^{\rm He}_{\rm esc}$ = 0.025. Upper
and lower panels display $F_{\rm HII}$ and $F_{\rm HeIII}$, with solid and
dashed lines in each panel representing continuous and bursty star
formation respectively with evolving Pop III spectra.}
\end{figure}

\clearpage

\begin{figure}
\includegraphics[height=3.5in]{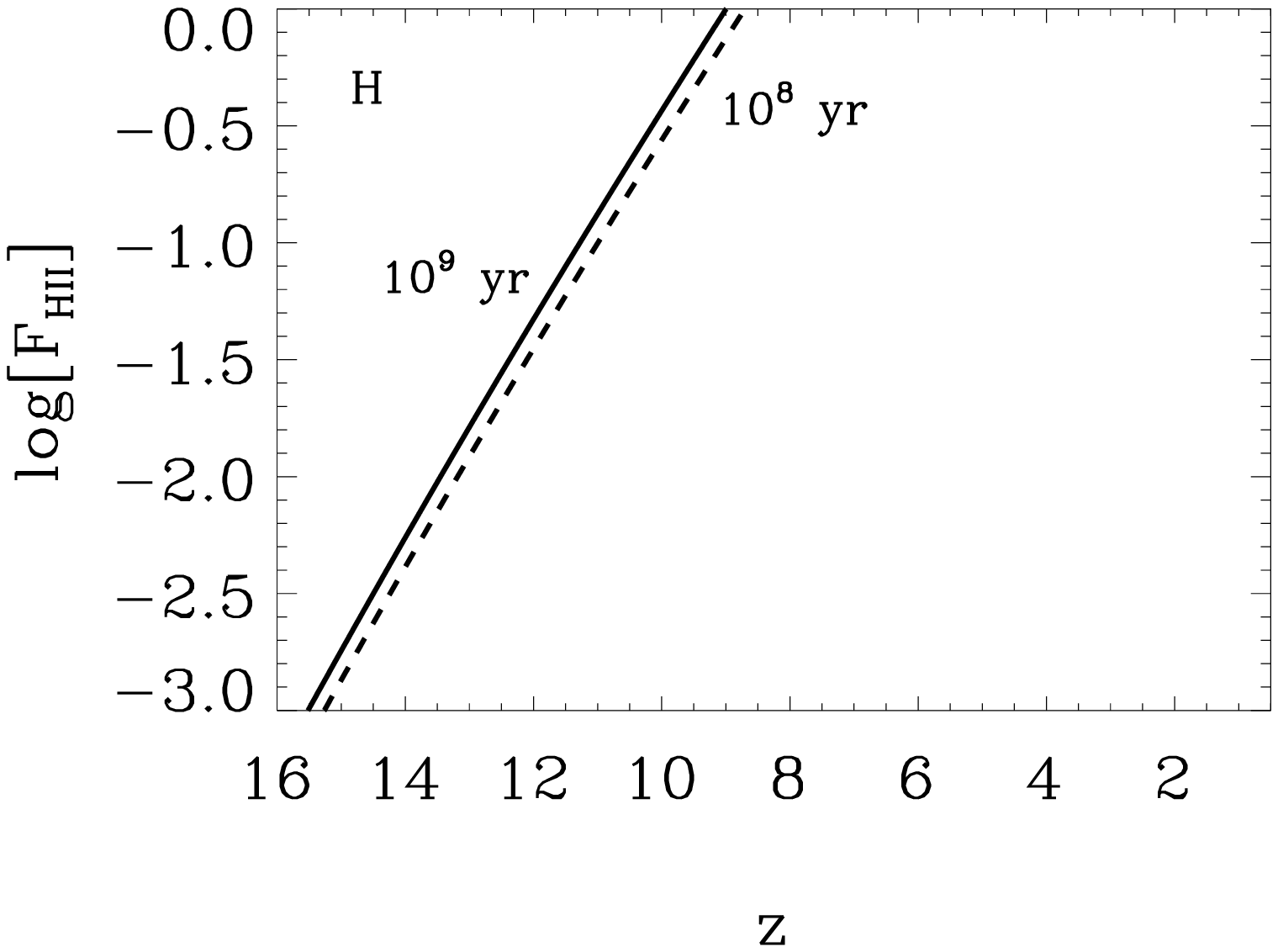}
\includegraphics[height=3.5in]{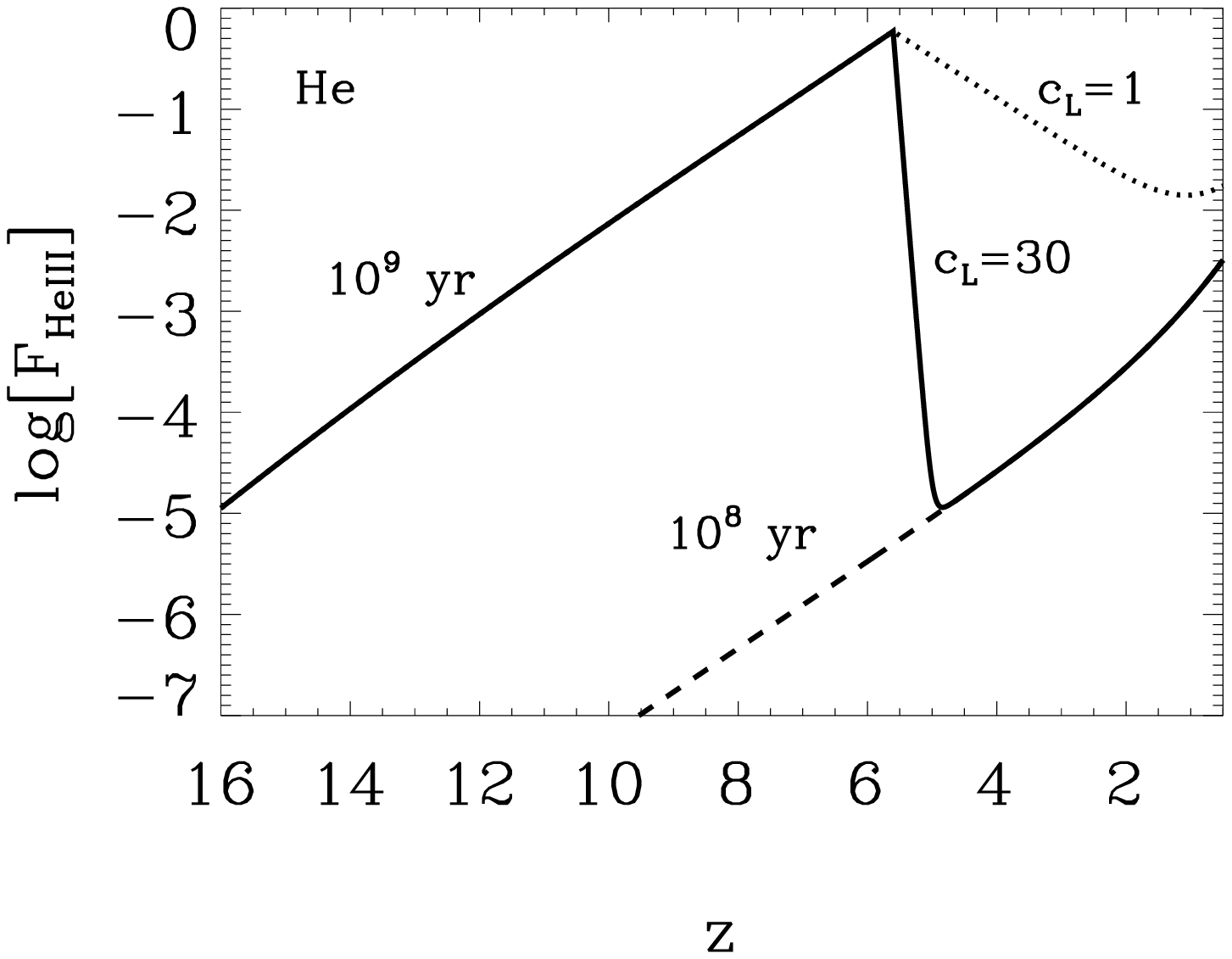}
\caption{The redshift evolution of the volume filling factors of H~II,
$F_{\rm HII}$, and He~III, $F_{\rm HeIII}$, are shown for the standard
model in this work. Upper and lower panels display $F_{\rm HII}$ and
$F_{\rm HeIII}$ respectively for continuous star formation. Solid and
dashed lines in each panel represent Pop III spectra switching to Pop II at
times corresponding to ages of the universe of $10^9$ and $10^8$ yr. For
$F_{\rm HeIII}$, an additional case is shown with a dotted line, where Pop
III spectra switch to Pop II at $10^9$ yr with the assumption of a
homogeneous IGM ($c_L$ = 1) at $z \la$ 5.6. This is intended to mimic the
fate of low-density regions in the IGM subsequent to partial ionization
(see text).}
\end{figure}

\clearpage

\begin{figure}
\plotone{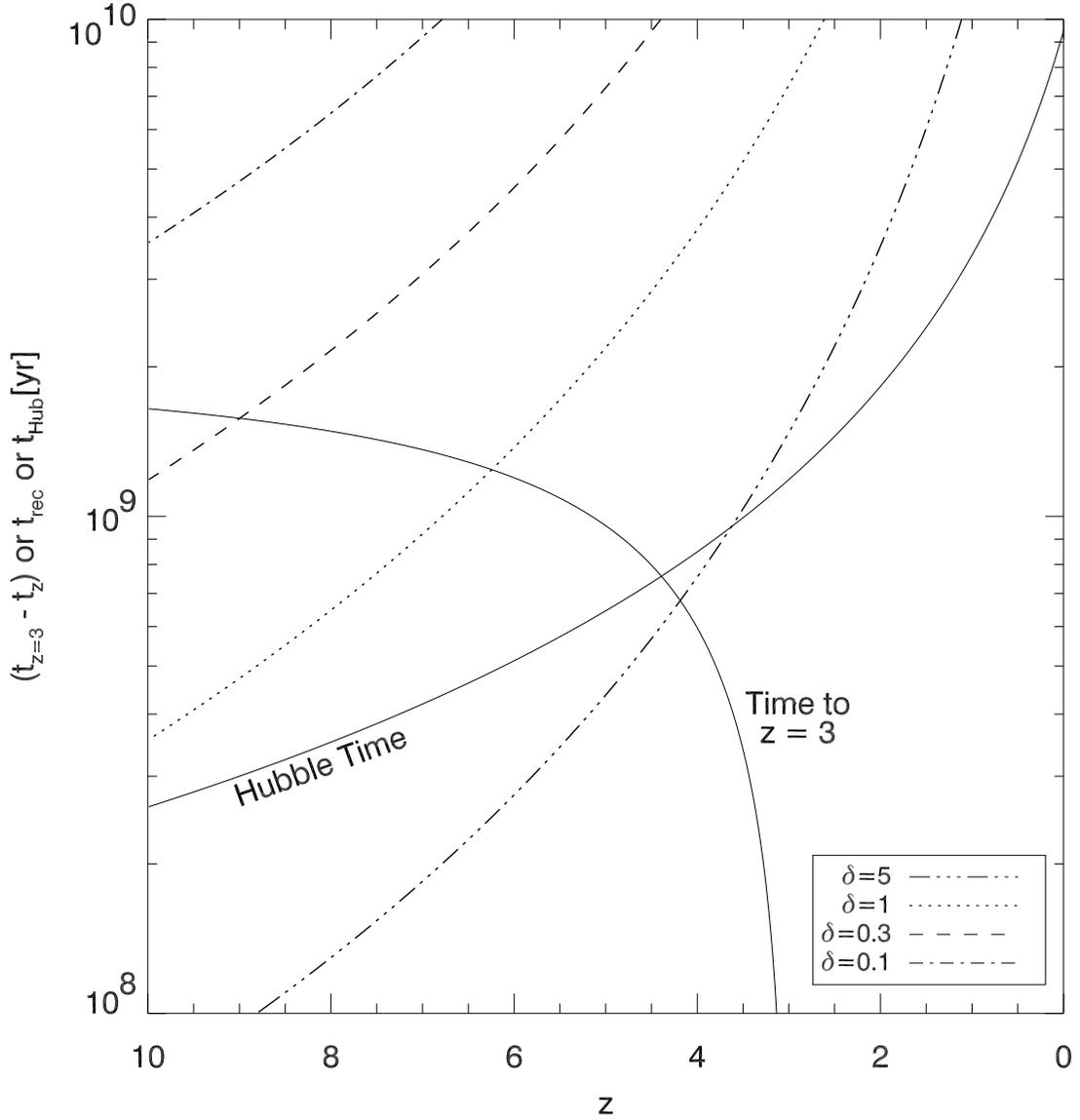}
\caption{Comparison of four He~III to He~II recombination times (e-folding
timescales; see text) for four values of overdensity $\delta$ in the IGM
with the elapsed time from redshift $z$ to $z = 3$, as a function of
redshift. For $\delta \lesssim 1.0$, the recombination time is longer than
the time to $z = 3$ and the Hubble time. These regions will not have time
to recombine before $z = 3$, where we have observational constraints.  We
have neglected the photoionization from high-redshift quasars, whose
effects could be significant at $3 < z < 6$. \label{fig:recfig}}
\vspace{0.05in}
\end{figure}

\clearpage

\begin{table}
\renewcommand{\arraystretch}{1.2}
\begin{center}
\caption{Evolution of Baryons in Halos and IGM Ionization Fractions}
\vspace{0.4in}
\begin{tabular}{cccccccccc}
\hline \hline
 $z$  & $F_{\rm B}$ & \multicolumn{4}{c}{$F_{\rm HII}$} &
\multicolumn{4}{c}{$F_{\rm HeIII}$} \\
 & & &    &  & &  & \\
\hline
& & Pop III & Pop III  & \multicolumn{2}{c}{Pop III/II}  & Pop III  & Pop III 
&   \multicolumn{2}{c}{Pop III/II} \\
& & (bursty)& (cont.)  & $10^8$ yr & $10^9$ yr & (bursty) & (cont.)  & $10^8$
yr & $10^9$ yr   \\
& & &  &  & &  & \\
20 & $2.7_{-7}$ & $2.1_{-6}$ & $4.4_{-6}$ & $3.3_{-6}$ & $4.4_{-6}$ &
$1.3_{-8}$ & $8.5_{-8}$ & $7.2_{-13}$ &$8.5_{-8}$ \\
15 & $4.9_{-5}$ & $4.6_{-4}$ & $1.8_{-3}$ & $1.3_{-3}$ & $1.8_{-3}$ &
$2.5_{-6}$ & $3.5_{-5}$ & $3.0_{-10}$ &$3.5_{-5}$ \\
10 & $3.3_{-3}$ & $3.4_{-2}$ & 0.37       & 0.27  & 0.37     & $1.7_{-4}$ &
$7.4_{-3}$          & $6.2_{-8}$  &$7.4_{-3}$ \\
9  & $6.8_{-3}$ & $7.1_{-2}$ & $\sim$1   & 0.74  & $\sim$1   & $3.6_{-4}$ &
0.02   & $1.7_{-7}$  & 0.02     \\
7 & $2.6_{-2}$ & 0.27 & $\sim$1 & $\sim$1 & $\sim$1 & $1.6_{-3}$ & 0.15 & $1.2_{-6}$ & 0.15 \\
6  & $4.7_{-2}$ & 0.53  & $\sim$1  & $\sim$1 & $\sim$1 & $2.9_{-3}$ & 0.39  & $3.3_{-6}$  &
 0.39 \\ 
4 & 0.13 & $\sim$1 & $\sim$1 & $\sim$1 & $\sim$1 & $1.0_{-2}$ & $\sim$1 & $2.6_{-5}$ & $2.6_{-5}$ \\
3  & 0.22  & $\sim$1 & $\sim$1  & $\sim$1  & $\sim$1 & $3.2_{-2}$ & $\sim$1 & $7.9_{-5}$ &$7.9_{-5}$ \\
2 & 0.33 & $\sim$1 & $\sim$1 & $\sim$1 & $\sim$1 & 0.12 & $\sim$1 & $2.8_{-4}$ & $2.8_{-4}$ \\
0  & 0.62 & $\sim$1  & $\sim$1 & $\sim$1 & $\sim$1 & $\sim$1 & $\sim$1  & $8.9_{-3}$ & $8.9_{-3}$  \\
 & & &  &  &   &  & \\
$z_{\rm reion}$ & & 4.7 & 9  & 8.7 & 9 & 0.7 & 5.1 & ... & ... \\
 & & &  &  &  & & & & \\
\hline
\end{tabular}
 
\tablecomments{The redshift evolution of the collapsed baryon fraction,
$F_{\rm B}$, and the volume filling factors $F_{\rm HII}$ and $F_{\rm
HeIII}$ (or equivalently, the volume-averaged IGM H~II and He~III
ionization fractions), for the four standard model scenarios (see text)
displayed in Figures 2--3, with star formation beginning at $z = 25$. For
the cases in the last two columns, He~II never reionizes completely, but
reaches a maximum ionization fraction of $\sim$ $8.9_{-3}$ (0.6) at $z \sim
0$ (5.6) for the duration of Pop III star formation being $10^8$ ($10^9$)
yr.  In our notation, $1.0_{-3} \equiv 1.0\times 10^{-3}$.}

\end{center}
\end{table}

\clearpage

\begin{table}
\renewcommand{\arraystretch}{1.2}
\begin{center}
\caption{Reionization Epoch and Small-Scale Power}
\vspace{0.4in}
\begin{tabular}{lcccc}
\tableline \tableline
 & & \multicolumn{3}{c}{$z_{\rm reion, H}$} \\ 
 & & \multicolumn{3}{c}{$\sigma_8$ = 0.7 ($\sigma_8$ = 0.9)} \\ \tableline
  & & $n = 0.9$ & $n = 1$  & $n = 1.1$ \\ \tableline
Pop III, bursty & & 3.5 (4.2) & 3.9 (4.7) & 4.4 (5.4) \\
Pop III, continuous & & 5.9 ({\bf 7.7}) & {\bf 7.0} ({\bf 9.0})  & {\bf 8.3} (10.6) \\
\tableline
\end{tabular}

\tablecomments{The redshifts, $z_{\rm reion, H}$, of hydrogen reionization
for $\sigma_8$ = 0.7 and $\sigma_8$ = 0.9, and for values of $n$ spanning
its current range from observations (0.9 -- 1.1). All other parameters are
fixed at their standard model values. The two modes of Pop III star
formation indicate the possible ranges for $z_{\rm reion, H}$ for a given
set of parameters; in particular, the continuous star formation case
represents the most extreme effects of metal-free stars.  The cases that
are consistent with current data on $z_{\rm reion, H}$ are shown in bold
type.}

\end{center}
\end{table}

\end{document}